\documentclass[twocolumn,secnumarabic,amssymb, nobibnotes, aps, prl,fleqn,amsmath,mathtools, nofootinbib,pgfmath,pgffor,superscriptaddress]{revtex4-1}
\usepackage{bm}        % for math
\usepackage{amssymb}   % for math
\usepackage{amsmath}   % for math
\usepackage{graphicx}% Include figure files

\makeatletter
\def\qrr@split@result#1 #2\@qrr@split@result{\edef\erfInput{#1}\edef\erfResult{#2}}
\newcommand*{\gnuplotErf}[2][\jobname.eval]{%
    \immediate\write18{gnuplot -e "set print '#1'; print #2, erf(#2);"}%
    \everyeof{\noexpand}
    \edef\qrr@temp{\input{#1} }%
    \expandafter\qrr@split@result\qrr@temp\@qrr@split@result
}
\makeatother

\usepackage{caption,color}
\usepackage[caption=false]{subfig}
\usepackage{multirow}

\usepackage{hyperref}
\hypersetup{colorlinks,citecolor=blue}

\usepackage[dvipsnames]{xcolor}

\begin{document}

%\title{\textcolor{blue}{Gravitational microlensing of gravitational waves from binary Primordial Black Holes }}
%\title{ Examining the source of gravitational wave events in the advanced LIGO detector}
%\title{ Primordial Black Holes as the gravitational wave source of advanced LIGO events}
\title{ Possibility of primordial black holes as the source of gravitational wave events in the advanced LIGO detector}

%\input author_list.tex       % D0 authors (remove the first 3 lines
                             % of this file prior to submission, they
                             % contain a time stamp for the authorlist)
                             % (includes institutions and visitors)
%\date{\today}

\author{E. Khalouei}
\author{H. Ghodsi}
\author{S. Rahvar}
%\affiliation{Department of Physics, Sharif University of Technology, 113659161, Tehran, Iran}
\affiliation{Department of Physics, Sharif University of Technology, P.O. Box 11365-9161, Tehran, Iran}
\author{J. Abedi}
\affiliation{Max-Planck-Institut f\"ur Gravitationsphysik, D-30167 Hannover, Germany}
\affiliation{Leibniz University\"at Hannover, D-30167 Hannover, Germany}
%$^{2}$ .............?}

%\author{Jahed Abedi}
%\email{jahed.abedi@aei.mpg.de}

\begin{abstract}
The analysis of gravitational Wave (GW) data from advanced LIGO provides the mass of 
each companion of binary black holes as the source of GWs. The mass of events corresponding to the binary black holes from GW is above $20$  M$_\odot$ which is much larger than the mass of astrophysical black holes detected by x-ray observations. In this work, we examine primordial black holes (PBHs) as the source of LIGO events. Assuming that $100\%$ of the dark matter is made of PBHs, 
we estimate the rate at which these objects make binaries, merge, and produce GWs as a function of redshift. The gravitational lensing of GWs by PBHs can also enhance the amplitude of the strain. We simulate GWs sourced by binary PBHs, with the detection threshold of $S/N>10$  for both Livingston and Hanford detectors. For the log-normal mass function of PBHs, we 
generate the expected distribution of events,  compare our results with the observed events,
 and find the best value of the mass function parameters (i.e.,  $M_c =25 M_\odot$ and $
\sigma=0.6$) in the log-normal mass function. Comparing the expected number of events with the number of observed ones rules out the present-Universe binary formation PBH scenario as the candidate for the source of GW events detected by  LIGO.

\end{abstract}

%\pacs{}

\maketitle

The recent discovery of gravitational waves (GWs) by advanced LIGO, sourced from  
binary black hole mergers, opened a new window in astronomy. The analysis of strain 
signals with theoretical templates of the GW can determine the mass of binary black holes (BBHs)
\cite{Abbott}. Since the mass of black holes from GWs are larger than the astrophysical black holes 
discovered yet by x-ray observation \cite{Casares}, one of the possibilities could be that the source of observed GWs is due to 
primordial black holes (PBHs).  
Ten GW candidates are detected as a result of the first and second runs of advanced 
LIGO (i.e., O1 and O2) \cite{Abbott} and VIRGO \cite{Asc} in $48.6$ day and $117$ day runs, 
respectively \cite{dayo1,dayo2}. %(where other searches for sub-solar mass black holes are not found in data \cite{Nitz:2020bdb}).
During the third observing run $O_{3}$ (i.e., April $2019$-March $2020$)\cite{oii} with the improved sensitivity of detectors \cite{Acernese,Tse}, ten candidate GW events have been identified \cite{oi}; however, four events have been confirmed \cite{a1,a2,a3,a4}.
 %\textcolor{red}{From the last statement it is  not clear either the number of events are few tens of events or 4 event ?}.

%??ina serch konam...$cite 200514016$}

% has revived interest in studying primordial black holes (PBHs) as  being dark matter candidates. 
As pointed in Refs. \cite{Carr,Carr2020}, and references therein, PBHs cannot make all the dark matter regarding the monochromatic mass function. Several scenarios to describe the formation of PBHs based on the collapse of large density perturbation in the radiation-dominated era propose an extended mass function for PBHs \cite{Kuhnel,Niemeyer}. That would make possible that PBHs with an extended mass function make all dark matter of the Universe\cite{Carr,Carr2020,Green}.

%{\bf While most constraints the PBHs assume they are produced with a monochromatic mass function \cite{kepler,gamma,widebinary,GC}, PBH formation demands them to follow an extended mass function \cite{Kuhnel,Carr,Carr2020}.
%the PBHs assume they are produced with a monochromatic mass function, 
%Moreover, taking the more extended mass function for the PBHs increases the 
%chance that they form $100\%$ of the dark matter \cite{Carr,Green}.}
%investigate the PBH mass function and compare a few models against each other and 
%they find parameter combinations for each distribution where the PBHs can make up 
%all the dark matter. 
%Being the formulation that fits a very large class of inflationary PBH models \cite{Kuhnel}, 
In this work, we adapt the log-normal mass function for the PBHs \cite{Carr,Kannike,Dolgov} as
\begin{equation}
\label{logn}
\psi(M)=\frac{f_{PBH}}{\sqrt{2 \pi} \sigma M} \exp\left(-\frac{log^{2}({M}/{M_{c}})}{2 \sigma^{2}}\right),
\end{equation}
where $M_{c}$ is characteristic mass, $\sigma$ is the width of mass function, and $f_{PBH}$ is the
fraction of dark matter made of PBHs.
%The three mass windows that Carr et al. \cite{Carr} demonstrate could make up the whole of dark matter centre around $10^{-16}$, $10^{-14}$ and $10^2$ solar masses.

While PBHs in the early universe were produced individually, we can argue that a very small fraction of
them can make binary systems in a cosmological timescale. When two individual PBHs pass close to each other, as a result of gravitational interaction, they can radiate GWs. For efficient interaction,  PBHs can make binary systems due to the dissipation process of GW emission surpassing their initial kinetic energy \cite{Bird}. The rate of binary formation per halo is given by \cite{Bird}
\begin{equation}
R(M)  =  \int_{0}^{R_{\rm vir}} 2 \pi r^2  \Big( \frac{\rho_{\rm nfw}(r)}{M_{\rm pbh}} \Big)^2 \langle \sigma_{GW} v_{\rm pbh} \rangle dr,
\label{rate0}
\end{equation}
%where 
%{\bf Write the cross section for GW radiation}
where $\sigma=\pi (\frac{85 \pi}{3})^{2/7} R_{\rm s}^{2} (\frac{v_{\rm pbh}}{c})^{18/7}$ is the cross section for binary formation \cite{cross1,cross2}.
$\rho_{\rm nfw} = \rho_s[(r/R_s)(1+r/R_s)^2]^{-1}$ is the Navarro-Frenk-White (NFW) density profile with characteristic radius and density $R_s$ and $\rho_s$, respectively. $R_{\rm vir}$ is the virial radius of the galactic halo at which the density of NFW profile reaches 200 times the mean cosmic density. $v_{\rm pbh}$ is the dispersion velocity of PBHs, and $M_{\rm pbh}$ is the mass of the PBH.
%To get a rough estimate of the fraction of binary formation per halo, assume a dark matter halo with mass $M=10^{12}  M_\odot$. The merger rate for the halo is the order of $10^{-13}$ year$^{-1}$ (see Figure $(1)$ of \cite{Bird}). The probability of binary formation during age of universe would be $\sim 10^{10}/10^{13}= 10^{-3}$ for an halo with mass $M=10^{12}  M_\odot$. If we have a sample of $1000$ PBHs, just one binary system could been produced.}

\begin{figure}
	\includegraphics[angle=0,width=0.49\textwidth,clip=]{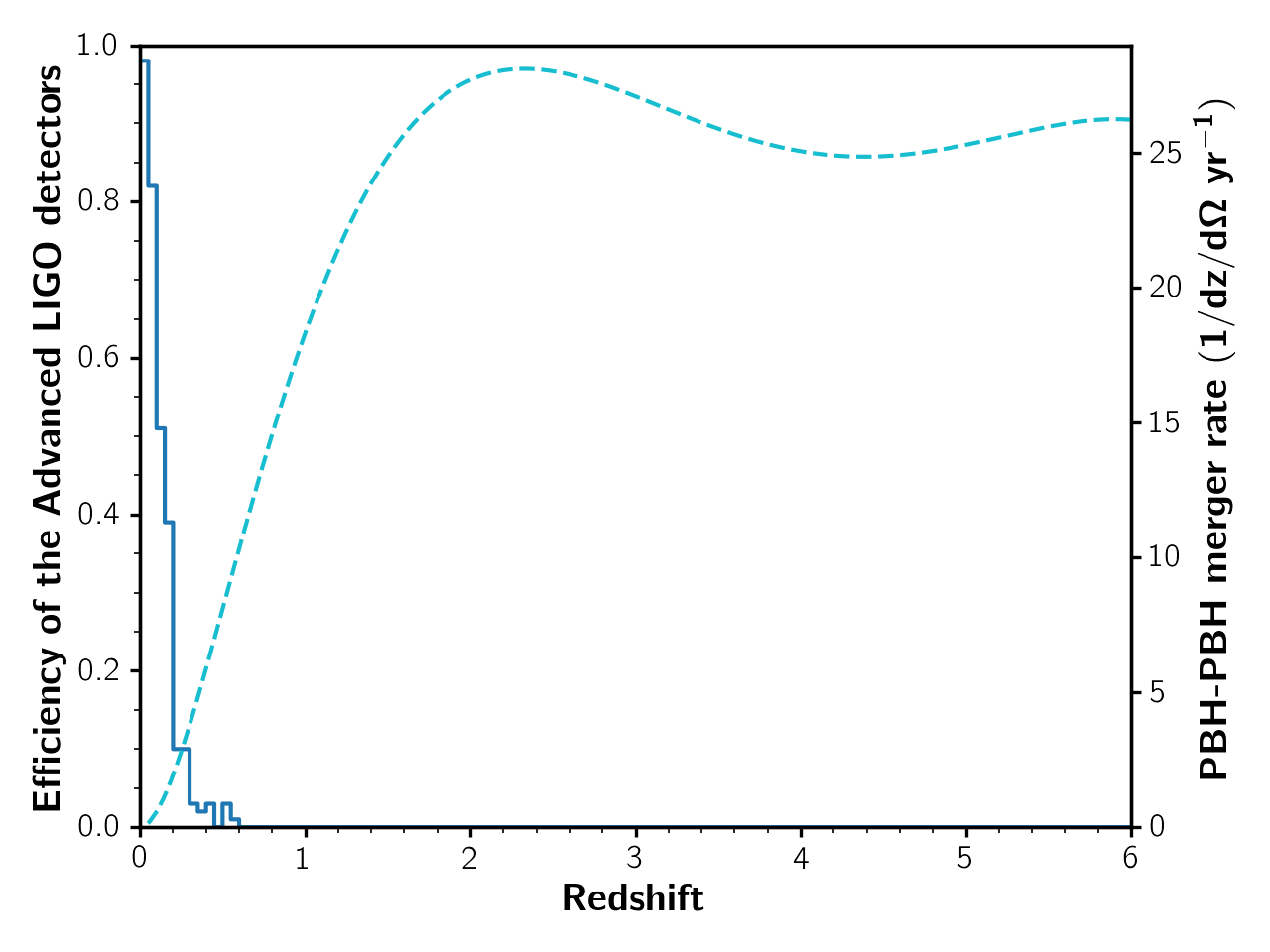}
	\caption{The dashed line represents the total PBH merger rate per steradian per redshift per year from Eq. (\ref{gamma}). Here, we assume that 100\% of dark matter is made of PBHs.  The solid-line histogram shows the redshift dependence of the efficiency of advanced LIGO with $SNR>10$ detection threshold.
 }
	\label{rate}
\end{figure}

After the binary formation stage, the binary system can further emit GWs, lose orbital energy, and finally inspiral to merge. This merging happens on a timescale which is much smaller than the Hubble time at $z=0$, and, hence, at the current time, we can safely assume that once a binary is formed, it will certainly also merge to produce GWs \cite{Bird}. In order to estimate how far back we can go in time to have the first merging, we compare the merging timescale \cite{OLeary} with the age of the Universe [i.e., ${4\pi G M_{tot}}/{v_{pbh}^{3}}\sim 1/H(z)$]. % \textcolor{red}{check the details of this equation}.  
  Numerical estimation from this equation results in 
 %  we set these two times equal and arrive at a redshift of
   $z\approx6$. 
   
   %That is we set equal the age of the Universe, $1/H(z)$, with the merging timescale, $\frac{4\pi G M_{tot}}{w^{3}}\sim \frac{1}{H(z)} $. Here $ M_{tot}$ is the total mass of PBHs. $G$, $w$ and $H(z)$ are the gravitational constant, the initial relative velocity of PBHs and Hubble time respectively. We therefore place a cut-off at this redshift for our calculations.

In what follows, we estimate the number of binary PBHs that can be formed from single PBHs. We follow the work of Bird et al. \cite{Bird}, where the calculation is done for the local Universe. 
The total merger rate per unit volume is as follows:
\begin{equation}
\label{rate2}
\Gamma = \int \frac{dn}{dM} R(M) dM,
\end{equation}
where $dn/dM$ is the halo mass function that we adopt the Press-Schechter formalism \cite{Press} and $R(M)$ from Eq. (\ref{rate0}) is the rate of mergers in a given halo of mass $M$. 

%Here, for the mass function we  and for the rate relation, we use the estimate given by Bird et. al. \cite{Bird} which utitlizes the NFW galaxy density profile. The rate per halo can be computed using: 

%\begin{equation}
%R(M)  = 4 \pi \int_{0}^{R_{\rm vir}} r^2 \frac{1}{2} \Big( \frac{\rho_{\rm nfw}(r)}{M_{\rm pbh}} \Big)^2 \langle \sigma v_{\rm pbh} \rangle dr,
%\end{equation}

Considering that halos are formed at the higher redshifts \cite{Ludlow}, we take into account the redshift dependence in Eq. (\ref{rate2}) and write the rate of events in comoving volume as
\begin{equation}
\frac{dN(z)}{dzd\Omega} =  c\frac{\chi^{2}(z)}{H(z)} \int_0^{M_{vir}(z)}  \frac{dn(z)}{dM} R(M,z)dM,
\label{gamma}
\end{equation}
where  $c$ is the speed of light, $\chi(z)$ is the comoving distance, $H(z)$ is the Hubble expansion rate, and $dn(z)/dM$ is the halo mass function of halos at the redshift $z$. It should be noted that the upper limit of the integral decreases with increasing the redshift where at the higher redshifts the larger-mass halos have not  been virialized yet \cite{Loeb}. 

From Eq. (\ref{gamma}), we plot $dN(z)/{dzd\Omega}$ in Fig. \ref{rate} (dashed line) as the merging rate per redshift per steradian per year, assuming that $100\%$ of dark matter is made of PHBs. It is noteworthy that the peak at $z\approx2$ results from the peaking comoving volume at this redshift. Integrating the curve in Fig. \ref{rate} over the redshift, the result is about $1724$ events per year for $z<6$. 
However, as we will determine the detection efficiency of LIGO, only the close-by events can be detected. GW signals for distant sources can also be magnified by gravitational lensing.  Assuming the 
whole dark matter is made of PBHs, they can also play the role of lensing. 

In what follows, we simulate the GWs from PBHs binary merging in the Universe and measure the detectability of events  by LIGO detector.  For simulating an ensemble of GW sources and lenses, we assume (a)  a mass function for PBHs is log normal \cite{Carr} and (b) the redshift distribution of the PBHs follows the comoving volume. Then we calculate the antenna pattern function for Hanford and Livingston and include the background noise to determine GW strain in the detector frame \cite{Abbott,pyc}. 

\begin{figure}
%	\centering
	\includegraphics[angle=0,width=0.49\textwidth,clip=]{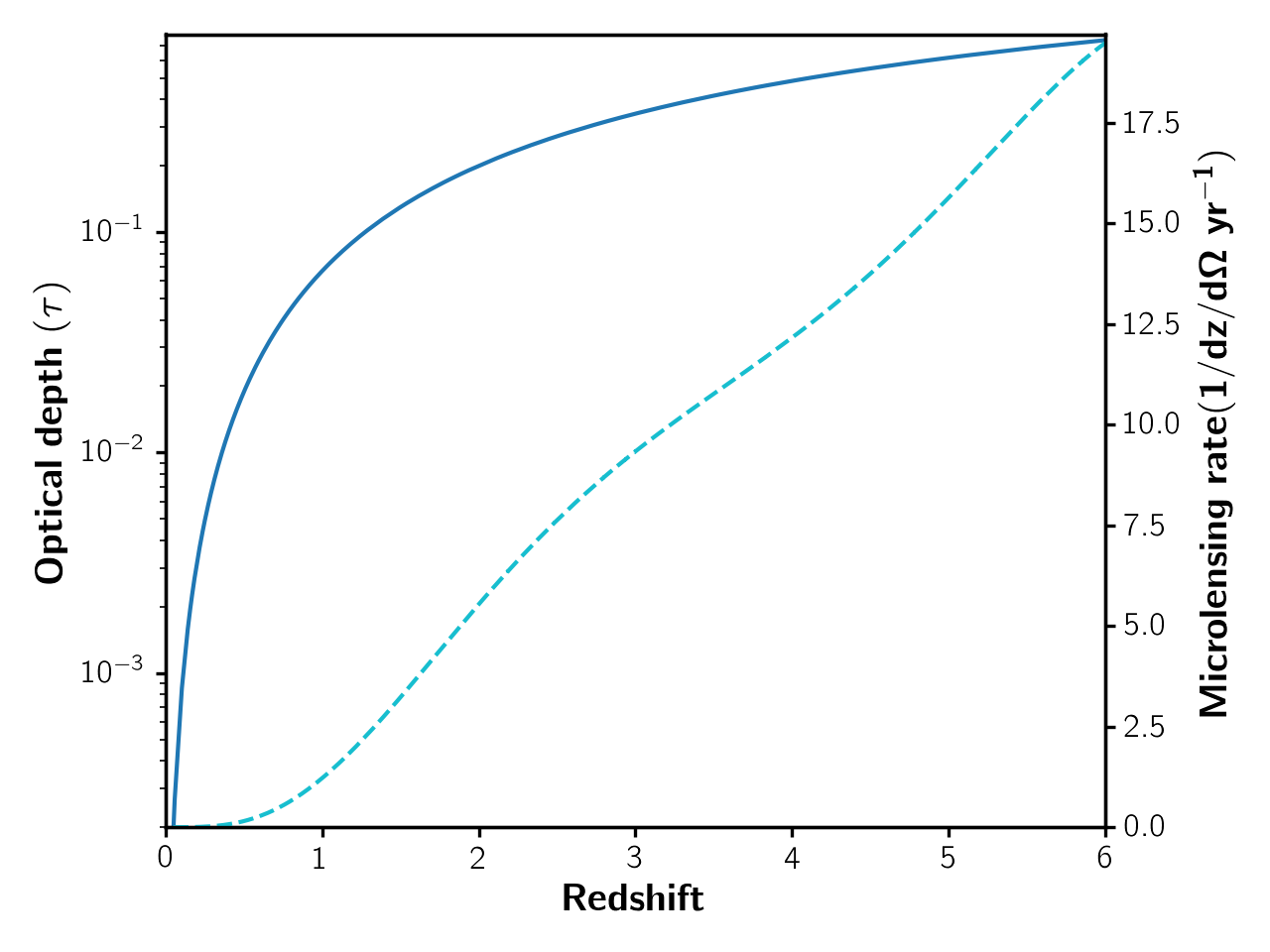}
	\caption{This plot assumes that the whole of dark matter is composed of PBHs(light blue). The multiplication of the optical depth $\tau(z)$ (dark blue curve) and the event rate, $N(z)$ as a function of the redshift. This curve represents the rate of microlensed GWs, sourced by PBHs.}\label{tau}
\end{figure}

In the next step, we calculate the microlensing optical depth as a function of redshift to calculate the number of GWs being lensed. The optical depth as the probability of lensing for a homogeneous distribution of PBHs is give by \cite{Zackrisson}
\begin{equation}
\tau = \frac{3H_0\Omega_{\rm PBH}}{2D_{\rm s}} \int_0^{z_{\rm s}} \frac{(1+z)^2D_{\rm ls}D_{\rm l} \rm dz}{c\sqrt{\Omega_{\rm M}(1+z)^3 + \Omega_{\rm \Lambda}}},
\end{equation}
where $H_0$ is the current value of the Hubble parameter, $\Omega_{\rm M}$ is the dark matter density parameter, and $\Omega_{\rm \Lambda}$ is the dark energy density parameter. $D_{\rm s}$, $D_{\rm ls}$, and $D_{\rm l}$ are the comoving observer-source distance, lens-source distance, and lens distance, respectively. Here, we assume PBHs composed the whole dark matter (i.e., $\Omega_{\rm M}=\Omega_{\rm PBH}$). Figure \ref{tau} (solid  curve) shows the optical depth as a function of redshift. The multiplication of the optical depth, $\tau(z)$ by the event rate $N(z)$ would result in the rate of the microlensed-GW event. This rate is shown in Fig. \ref{tau} (dashed curve).

%%%%%%%%%%%%%%%%%%%%%
%%%%%%% STRAIN %%%%%%%%%
%%%%%%%%%%%%%%%%%%%%%

%We assume the binary black holes as the source of GWs (during the spiral, merger and ring down stages). 
%\textcolor{blue}{The waveform model of GWs can be calculated from numerical simulation (IMRPhenomPv$2$ model) \cite{wavef} using the PyCBC package \cite{pyc}}.
%The GWs of non-spinning binaries in the frequency space from the numerical solution of Einstein equation (IMRPhenomPv$2$ model) \cite{pycbc} can be written as
%\begin{equation}
%h(f)=A_{eff}(f) e^{i \psi_{eff}(f)}
%\end{equation}
%where $A_{eff}(f)$ is the effective amplitude which depends on the comoving distance to the source from the observer $D(z)$. Also it depends on the angular location of PBHs binary in sky. $\psi_{eff}(f) $ is the phase which depends on the total mass and symmetric mass ratio $\eta ={m_{1}m_{2}}/{M_{total}^{2}}$ of PBHs.

%\magenta{note on 100 solar mass?}

%where $\epsilon$ is the efficiency of energy release, $M_{\rm tot}$ is the total mass involved and $D(z)$ is the comoving distance to the source from the observer. 

%%%%%%%%%%%%%%%%%%%%%%%%
%%%% MICROLENSING RATE %%%%%%%%%
%%%%%%%%%%%%%%%%%%%%%%%%

In the gravitational lensing of gravitational waves, since the mass of the binary merging system is on the order of the
lens, then the wavelength of GW $\lambda$  is on the order of the 
Schwarzchild radius of the lens, $R_{sch}$. In this case, we need to deal with the problem with the wave optics approach. 
The propagation of gravitational wave  with a perturbation due to a lens behaves similar to the electromagnetic equation (i.e. $\square h^{\mu\nu} = 0$) \cite{De,Takahashi}, where the metric is $g_{\mu\nu} = \eta_{\mu\nu} + h_{\mu\nu}$. The GWs can be magnified similar to the electromagnetic radiation during the microlensing; however, since the timescale of GW is very short, we can take a static configuration for the relative position of the source, lens and the observer.

From the generic solution of the wave equation, we can calculate the magnification factor for the strain of the GW during the lensing as \cite{Schneider}
\begin{equation}
\mu_{GW}(\beta;k) = \sqrt{\pi f}J_{0}(f\beta),
\label{mu}
\end{equation}
where $f = 2 k R_{\rm sch}$, $k$ is the wave number, $\beta$ is the impact parameter of the source on the lens plane normalized to the Einstein angle, and $J_0$ is the Bessel function of the first kind. We note that the magnification for light is the square of relation (\ref{mu}) as for the electromagnetic waves; we measure the intensity of light (i.e., $I\propto E^2$) while for the GW we  measure the strain (i.e., $h_{\mu\nu}$).   In the limit of $k \rightarrow \infty$, we can recover the geometric optics relation for the magnification. 
In our simulation, we limit the minimum impact parameter to be in the range of $0<\beta<1$.

In what follows, we consider 
log-normal mass functions for PBHs with different parameters of $M_{c}$ and $\sigma$ of Eq. (\ref{logn}). 
We perform a Monte Carlo simulation and generate binary PBHs over the redshift range of $(0-6)$. We also take into account the microlensing of the GWs with the magnification given by Eq. (\ref{mu}).

After generating GW events in our simulation, we add the corresponding noise to the data and use the specification of LIGO for calculating the detection efficiency of the LIGO detector. For analyzing simulated data, 
we use PyCBC inference \cite{Biwer:2018osg} package with dynamic nested sampling Markov chain Monte Carlo algorithm, dynesty \cite{10.1093/mnras/staa278}. It is based on sampling the likelihood function for a hypothesis that gives a measure of the existence of a signal in the data. The sampler performs the full Bayesian parameter estimation for each injection. Then we can obtain the maximum SNR recovered from injection. We assume a threshold of $S/N>10$ for criteria of significant detection. For each injection, we used the inbuilt injection creation of the PyCBC package \footnote{\href{https://www.atlas.aei.uni-hannover.de/work/ahnitz/projects/docs/pycbc/_gh-pages/latest/html/inference/examples/bbh.html?highlight=injection}{PyCBC inference documentation $>>$ Simulated BBH example $>>$ 1. Create the injection}} using the IMRPhenomPv2 waveform. We choose two detector networks (Hanford and Livingston) for this analysis. We run the pipeline for $16304\times4\times6$ [accounting for $M_{c}=20, 25, 30$, and  $35 M_{\odot}$ and $\sigma=0.1, 0.2, 0.3, 0.4, 0.5$, and $0.6$ of the model (\ref{logn})] injections using detector sensitivity during the O2 run. It is known that LIGO noise varies over long periods of time \cite{Martynov:2016fzi}; in order to model the detector sensitivity (accounting for non-Gaussian and/or nonstationary background noise), we made $32^{\rm{Hanford}}\times32^{\rm{Livingston}}=1024$ number of power spectral densities (PSD) of 4096 s random data and perform the PyCBC pipeline to analyse the injections for these random times. We used  a fake Gaussian noise (via the fake-strain option) that is colored by a given PSD.
%We also used an analytical model of the advanced LIGO sensitivity curve named Zero-detuned-high power (ZDHP) noise curve\footnote{\href{https://dcc.ligo.org/LIGO-T1800044/public}{https://dcc.ligo.org/LIGO-T1800044/public}} to run the same simulation. In both cases, a fake Gaussian noise (via the fake-strain option) is generated that is colored by either a given PSD or the Advanced LIGO updated design sensitivity curve. Two results are comparable given that the design sensitivity is $\sim3$ times more sensitive than what the LIGO detectors were when GW150914 was detected. 

These injections are created with zero spin BBH components\footnote{Although it would be more physical to set a nonzero distribution for each component spin, we assume that it does not affect the signal recovery of injections.}. 
%We set a uniform in magnitude and isotropic in orientation prior on each component object’s spin in our search.
The orientation (inclination and polarization angles ) and location (right ascension and declination) of the GW sources are distributed uniformly in the polarization sphere and in the sky, respectively \cite{Abbott,india}.

\begin{figure}
	\includegraphics[angle=0,width=0.49\textwidth,clip=]{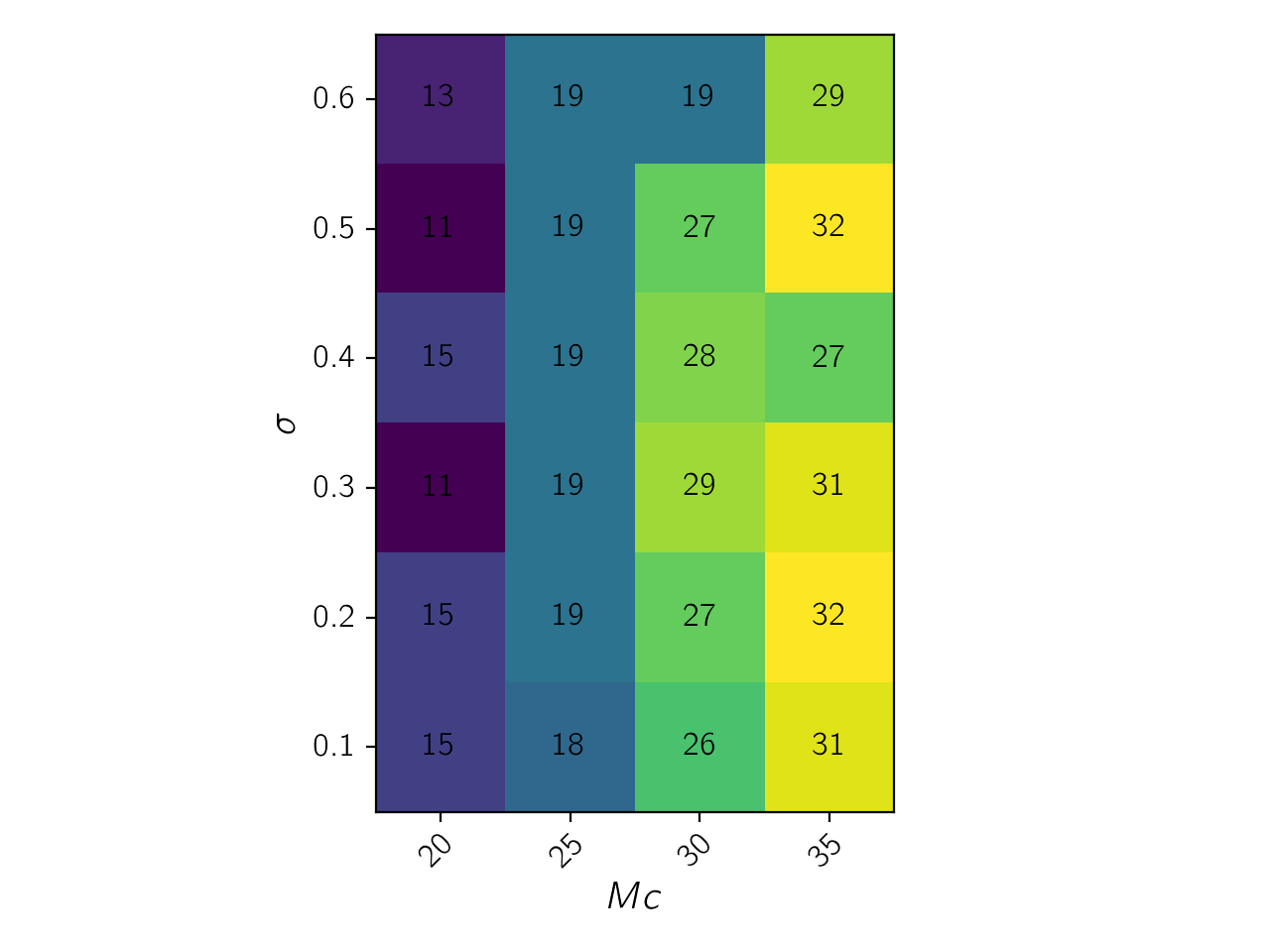}
\caption{
 The number of detected events from the simulated events for source located at $z<1$, for various log-normal parameters of $\sigma$ and $M_{c}$ PBH mass function, assuming $f=1$.}
	\label{rate2}
\end{figure}

In the following, we calculate the redshift-dependent detection efficiency function of advanced LIGO observatory as
\begin{equation}
\epsilon(z)=\frac{\Delta N_{detected}(z,z+\Delta z)}{\Delta N_{theory}(z,z+\Delta z)},
\end{equation}
where $\Delta N_{detected}$ is the number of events with $SNR>10$ between $(z,z+\Delta z)$ and $\Delta N_{theory}$ is the number of theoretical events we generate between $(z, z+\Delta z)$. 
In Fig. $\ref{rate}$, the histogram with a solid line represents the efficiency function of LIGO 
in terms of redshift. This function declines to zero at the redshift $z\simeq 0.6$. Multiplying the efficiency function by the normalized distribution of events based on the 
theoretical assumptions [i.e., $\Delta N_{expected} = \Delta N_{theory} \times \epsilon(z)$] results in the expected number of events.

In order to find the best parameters for the mass function of PBHs, we compare the mass distribution of observed GW sources with that of our simulated events \cite{Abbott}. Here, we use the mean and the width of the detected distribution of mass of events (derived from the analysis of the binary black hole) and compare them with the simulation. The log-normal mass function with a characteristic mass of $M_c=25 M_{\odot}$ and $\sigma = 0.6$ in Eq. (\ref{logn}) shown in  Fig. (\ref{rate2}) has the best compatibility between the theory and the observation.

 Now, we compare the number of observed events with that of our expectations from the theory. Integrating $\Delta N_{expected}/dz$ over $z$, we calculate the overall expected number of events, where, according to the efficiency function, we are able to detect events up to a distance of $z\sim 0.6$. We note that in our simulation $38\%$ of GW events with high redshifts have been microlensed; however, for the lower redshifts (i.e., $z<1$), the number of microlensed events is negligible. 
The comparison of the observed GW events ( seven events from 117 days observation by O2) with the expected number of events assuming $100\%$ of dark matter is made of PBHs (0.5 event for 117 days) reveals that $N_{observed} \gg N_{expected} $, which means that $f>1$.  In order to explain this result within the context of the scenario that has been used in this work, we propose the following two explanations. (i) The first possibility is that astrophysical black holes might be responsible for GW events instead of binary PBHs. The problem with this solution is that astrophysical black holes above $\sim$$20$ solar masses have not been discovered yet from direct x-ray observation. However, there are astrophysical scenarios for the formation of heavy black holes \cite{2010ApJ...714.1217B,2015MNRAS.451.4086S}.
 (ii) In this paper, we followed the formalism of PBH binary formation in the present Universe \cite{Bird}. 
There is other scenario that suggests the formation of a PBH binary in the early Universe \cite{Sasaki}. LIGO constraints the primordial black holes with a mass between $0.2$  M$_\odot$ $ and $ $1.0$  M$_\odot$  using  early-Universe PBH binary formation \cite{Abbottpbh,Abbottpbh2}. If we assume binary PBHs formation in early Universe, the number of binary black hole systems would increase and may explain the observed GW events.
\section*{ACKNOWLEDGMENTS}
The authors are grateful to Simeon Bird, Niayesh Afshordi and Mojahed Parsimood for their helpful comments and guidances. J. A. thanks Collin D. Capano, Sumit Kumar, and Alexander H. Nitz for answering his questions about PyCBC. Also we thank anonymous referee for useful comments improving this paper. This research was supported by Sharif University of Technology's Office of Vice
President for Research under Grant No. G950214. This work was supported by the Max Planck Gesellschaft, and we thank the Atlas cluster computing team at AEI Hanover. This research has made use of data, software, and/or web tools obtained from the Gravitational Wave Open Science Center (https://gw-openscience.org), a service of LIGO Laboratory, the LIGO Scientific Collaboration, and the Virgo Collaboration. LIGO is funded by the U.S. National Science Foundation. Virgo is funded by the French Centre National de Recherche Scientifique (CNRS), the Italian Istituto Nazionale della Fisica Nucleare (INFN), and the Dutch Nikhef, with contributions by Polish and Hungarian institutes. Furthermore, we acknowledge using the transfer function code of Eisenstein and Hu \cite{Hu}.

\bibliographystyle{apsrev}
\bibliography{gwfinal}

\providecommand{\noopsort}[1]{}\providecommand{\singleletter}[1]{#1}%
\begin{thebibliography}{42}
\expandafter\ifx\csname natexlab\endcsname\relax\def\natexlab#1{#1}\fi
\expandafter\ifx\csname bibnamefont\endcsname\relax
  \def\bibnamefont#1{#1}\fi
\expandafter\ifx\csname bibfnamefont\endcsname\relax
  \def\bibfnamefont#1{#1}\fi
\expandafter\ifx\csname citenamefont\endcsname\relax
  \def\citenamefont#1{#1}\fi
\expandafter\ifx\csname url\endcsname\relax
  \def\url#1{\texttt{#1}}\fi
\expandafter\ifx\csname urlprefix\endcsname\relax\def\urlprefix{URL }\fi
\providecommand{\bibinfo}[2]{#2}
\providecommand{\eprint}[2][]{\url{#2}}

\bibitem[{\citenamefont{Abbott et~al.}(2019{\natexlab{a}})}]{Abbott}
\bibinfo{author}{\bibfnamefont{B.}~\bibnamefont{Abbott}} \bibnamefont{et~al.}
  (\bibinfo{collaboration}{LIGO Scientific, Virgo}), \bibinfo{journal}{Phys.
  Rev. X} \textbf{\bibinfo{volume}{9}}, \bibinfo{pages}{031040}
  (\bibinfo{year}{2019}{\natexlab{a}}), \eprint{1811.12907}.

\bibitem[{\citenamefont{Casares et~al.}(2017)\citenamefont{Casares, Jonker, and
  Israelian}}]{Casares}
\bibinfo{author}{\bibfnamefont{J.}~\bibnamefont{Casares}},
  \bibinfo{author}{\bibfnamefont{P.~G.} \bibnamefont{Jonker}},
  \bibnamefont{and}
  \bibinfo{author}{\bibfnamefont{G.}~\bibnamefont{Israelian}},
  \emph{\bibinfo{title}{X-Ray Binaries}} (\bibinfo{publisher}{Springer
  International Publishing}, \bibinfo{address}{Cham}, \bibinfo{year}{2017}),
  pp. \bibinfo{pages}{1499--1526}, ISBN \bibinfo{isbn}{978-3-319-21846-5},
  \urlprefix\url{https://doi.org/10.1007/978-3-319-21846-5_111}.

\bibitem[{\citenamefont{Acernese et~al.}(2015)}]{Asc}
\bibinfo{author}{\bibfnamefont{F.}~\bibnamefont{Acernese}} \bibnamefont{et~al.}
  (\bibinfo{collaboration}{VIRGO}), \bibinfo{journal}{Class. Quant. Grav.}
  \textbf{\bibinfo{volume}{32}}, \bibinfo{pages}{024001}
  (\bibinfo{year}{2015}), \eprint{1408.3978}.

\bibitem[{\citenamefont{Abbott et~al.}(2016{\natexlab{a}})}]{dayo1}
\bibinfo{author}{\bibfnamefont{B.}~\bibnamefont{Abbott}} \bibnamefont{et~al.}
  (\bibinfo{collaboration}{LIGO Scientific, Virgo}), \bibinfo{journal}{Phys.
  Rev. X} \textbf{\bibinfo{volume}{6}}, \bibinfo{pages}{041015}
  (\bibinfo{year}{2016}{\natexlab{a}}), \bibinfo{note}{[Erratum: Phys.Rev.X 8,
  039903 (2018)]}, \eprint{1606.04856}.

\bibitem[{\citenamefont{Abbott et~al.}(2017)\citenamefont{Abbott, Abbott,
  Abbott, Acernese, Ackley, Adams, Adams, Addesso, Adhikari, Adya
  et~al.}}]{dayo2}
\bibinfo{author}{\bibfnamefont{B.~P.} \bibnamefont{Abbott}},
  \bibinfo{author}{\bibfnamefont{R.}~\bibnamefont{Abbott}},
  \bibinfo{author}{\bibfnamefont{T.}~\bibnamefont{Abbott}},
  \bibinfo{author}{\bibfnamefont{F.}~\bibnamefont{Acernese}},
  \bibinfo{author}{\bibfnamefont{K.}~\bibnamefont{Ackley}},
  \bibinfo{author}{\bibfnamefont{C.}~\bibnamefont{Adams}},
  \bibinfo{author}{\bibfnamefont{T.}~\bibnamefont{Adams}},
  \bibinfo{author}{\bibfnamefont{P.}~\bibnamefont{Addesso}},
  \bibinfo{author}{\bibfnamefont{R.}~\bibnamefont{Adhikari}},
  \bibinfo{author}{\bibfnamefont{V.}~\bibnamefont{Adya}}, \bibnamefont{et~al.},
  \bibinfo{journal}{Physical Review Letters} \textbf{\bibinfo{volume}{119}},
  \bibinfo{pages}{161101} (\bibinfo{year}{2017}).

\bibitem[{oii()}]{oii}
\bibinfo{howpublished}{\url{https://www.ligo.caltech.edu/news/ligo20200326}}.

\bibitem[{\citenamefont{Acernese et~al.}(2019)\citenamefont{Acernese, Agathos,
  Aiello, Allocca, Amato, Ansoldi, Antier, Ar\`ene, Arnaud, Ascenzi
  et~al.}}]{Acernese}
\bibinfo{author}{\bibfnamefont{F.}~\bibnamefont{Acernese}},
  \bibinfo{author}{\bibfnamefont{M.}~\bibnamefont{Agathos}},
  \bibinfo{author}{\bibfnamefont{L.}~\bibnamefont{Aiello}},
  \bibinfo{author}{\bibfnamefont{A.}~\bibnamefont{Allocca}},
  \bibinfo{author}{\bibfnamefont{A.}~\bibnamefont{Amato}},
  \bibinfo{author}{\bibfnamefont{S.}~\bibnamefont{Ansoldi}},
  \bibinfo{author}{\bibfnamefont{S.}~\bibnamefont{Antier}},
  \bibinfo{author}{\bibfnamefont{M.}~\bibnamefont{Ar\`ene}},
  \bibinfo{author}{\bibfnamefont{N.}~\bibnamefont{Arnaud}},
  \bibinfo{author}{\bibfnamefont{S.}~\bibnamefont{Ascenzi}},
  \bibnamefont{et~al.} (\bibinfo{collaboration}{Virgo Collaboration}),
  \bibinfo{journal}{Phys. Rev. Lett.} \textbf{\bibinfo{volume}{123}},
  \bibinfo{pages}{231108} (\bibinfo{year}{2019}),
  \urlprefix\url{https://link.aps.org/doi/10.1103/PhysRevLett.123.231108}.

\bibitem[{\citenamefont{Tse et~al.}(2019)\citenamefont{Tse, Yu, Kijbunchoo,
  Fernandez-Galiana, Dupej, Barsotti, Blair, Brown, Dwyer, Effler
  et~al.}}]{Tse}
\bibinfo{author}{\bibfnamefont{M.}~\bibnamefont{Tse}},
  \bibinfo{author}{\bibfnamefont{H.}~\bibnamefont{Yu}},
  \bibinfo{author}{\bibfnamefont{N.}~\bibnamefont{Kijbunchoo}},
  \bibinfo{author}{\bibfnamefont{A.}~\bibnamefont{Fernandez-Galiana}},
  \bibinfo{author}{\bibfnamefont{P.}~\bibnamefont{Dupej}},
  \bibinfo{author}{\bibfnamefont{L.}~\bibnamefont{Barsotti}},
  \bibinfo{author}{\bibfnamefont{C.~D.} \bibnamefont{Blair}},
  \bibinfo{author}{\bibfnamefont{D.~D.} \bibnamefont{Brown}},
  \bibinfo{author}{\bibfnamefont{S.~E.} \bibnamefont{Dwyer}},
  \bibinfo{author}{\bibfnamefont{A.}~\bibnamefont{Effler}},
  \bibnamefont{et~al.}, \bibinfo{journal}{Phys. Rev. Lett.}
  \textbf{\bibinfo{volume}{123}}, \bibinfo{pages}{231107}
  (\bibinfo{year}{2019}),
  \urlprefix\url{https://link.aps.org/doi/10.1103/PhysRevLett.123.231107}.

\bibitem[{oi()}]{oi}
\bibinfo{howpublished}{\url{https://gracedb.ligo.org/superevents/public/O3/}}.

\bibitem[{\citenamefont{Abbott et~al.}(2020{\natexlab{a}})\citenamefont{Abbott,
  Abbott, Abbott, Abraham, Acernese, Ackley, Adams, Adhikari, Adya, Affeldt
  et~al.}}]{a1}
\bibinfo{author}{\bibfnamefont{B.}~\bibnamefont{Abbott}},
  \bibinfo{author}{\bibfnamefont{R.}~\bibnamefont{Abbott}},
  \bibinfo{author}{\bibfnamefont{T.}~\bibnamefont{Abbott}},
  \bibinfo{author}{\bibfnamefont{S.}~\bibnamefont{Abraham}},
  \bibinfo{author}{\bibfnamefont{F.}~\bibnamefont{Acernese}},
  \bibinfo{author}{\bibfnamefont{K.}~\bibnamefont{Ackley}},
  \bibinfo{author}{\bibfnamefont{C.}~\bibnamefont{Adams}},
  \bibinfo{author}{\bibfnamefont{R.}~\bibnamefont{Adhikari}},
  \bibinfo{author}{\bibfnamefont{V.}~\bibnamefont{Adya}},
  \bibinfo{author}{\bibfnamefont{C.}~\bibnamefont{Affeldt}},
  \bibnamefont{et~al.}, \bibinfo{journal}{The Astrophysical Journal Letters}
  \textbf{\bibinfo{volume}{892}}, \bibinfo{pages}{L3}
  (\bibinfo{year}{2020}{\natexlab{a}}).

\bibitem[{\citenamefont{Abbott et~al.}(2020{\natexlab{b}})}]{a2}
\bibinfo{author}{\bibfnamefont{R.}~\bibnamefont{Abbott}} \bibnamefont{et~al.}
  (\bibinfo{collaboration}{LIGO Scientific, Virgo}), \bibinfo{journal}{Phys.
  Rev. D} \textbf{\bibinfo{volume}{102}}, \bibinfo{pages}{043015}
  (\bibinfo{year}{2020}{\natexlab{b}}), \eprint{2004.08342}.

\bibitem[{\citenamefont{Abbott et~al.}(2020{\natexlab{c}})\citenamefont{Abbott,
  Abbott, Abraham, Acernese, Ackley, Adams, Adhikari, Adya, Affeldt, Agathos
  et~al.}}]{a3}
\bibinfo{author}{\bibfnamefont{R.}~\bibnamefont{Abbott}},
  \bibinfo{author}{\bibfnamefont{T.}~\bibnamefont{Abbott}},
  \bibinfo{author}{\bibfnamefont{S.}~\bibnamefont{Abraham}},
  \bibinfo{author}{\bibfnamefont{F.}~\bibnamefont{Acernese}},
  \bibinfo{author}{\bibfnamefont{K.}~\bibnamefont{Ackley}},
  \bibinfo{author}{\bibfnamefont{C.}~\bibnamefont{Adams}},
  \bibinfo{author}{\bibfnamefont{R.}~\bibnamefont{Adhikari}},
  \bibinfo{author}{\bibfnamefont{V.}~\bibnamefont{Adya}},
  \bibinfo{author}{\bibfnamefont{C.}~\bibnamefont{Affeldt}},
  \bibinfo{author}{\bibfnamefont{M.}~\bibnamefont{Agathos}},
  \bibnamefont{et~al.}, \bibinfo{journal}{The Astrophysical Journal Letters}
  \textbf{\bibinfo{volume}{896}}, \bibinfo{pages}{L44}
  (\bibinfo{year}{2020}{\natexlab{c}}).

\bibitem[{\citenamefont{Abbott et~al.}(2020{\natexlab{d}})\citenamefont{Abbott,
  Abbott, Abraham, Acernese, Ackley, Adams, Adhikari, Adya, Affeldt, Agathos
  et~al.}}]{a4}
\bibinfo{author}{\bibfnamefont{R.}~\bibnamefont{Abbott}},
  \bibinfo{author}{\bibfnamefont{T.}~\bibnamefont{Abbott}},
  \bibinfo{author}{\bibfnamefont{S.}~\bibnamefont{Abraham}},
  \bibinfo{author}{\bibfnamefont{F.}~\bibnamefont{Acernese}},
  \bibinfo{author}{\bibfnamefont{K.}~\bibnamefont{Ackley}},
  \bibinfo{author}{\bibfnamefont{C.}~\bibnamefont{Adams}},
  \bibinfo{author}{\bibfnamefont{R.}~\bibnamefont{Adhikari}},
  \bibinfo{author}{\bibfnamefont{V.}~\bibnamefont{Adya}},
  \bibinfo{author}{\bibfnamefont{C.}~\bibnamefont{Affeldt}},
  \bibinfo{author}{\bibfnamefont{M.}~\bibnamefont{Agathos}},
  \bibnamefont{et~al.}, \bibinfo{journal}{Physical Review Letters}
  \textbf{\bibinfo{volume}{125}}, \bibinfo{pages}{101102}
  (\bibinfo{year}{2020}{\natexlab{d}}).

\bibitem[{\citenamefont{Carr et~al.}(2017)\citenamefont{Carr, Raidal, Tenkanen,
  Vaskonen, and Veerm\"ae}}]{Carr}
\bibinfo{author}{\bibfnamefont{B.}~\bibnamefont{Carr}},
  \bibinfo{author}{\bibfnamefont{M.}~\bibnamefont{Raidal}},
  \bibinfo{author}{\bibfnamefont{T.}~\bibnamefont{Tenkanen}},
  \bibinfo{author}{\bibfnamefont{V.}~\bibnamefont{Vaskonen}}, \bibnamefont{and}
  \bibinfo{author}{\bibfnamefont{H.}~\bibnamefont{Veerm\"ae}},
  \bibinfo{journal}{Phys. Rev. D} \textbf{\bibinfo{volume}{96}},
  \bibinfo{pages}{023514} (\bibinfo{year}{2017}), \eprint{1705.05567}.

\bibitem[{\citenamefont{Carr et~al.}(2020)\citenamefont{Carr, Kohri, Sendouda,
  and Yokoyama}}]{Carr2020}
\bibinfo{author}{\bibfnamefont{B.}~\bibnamefont{Carr}},
  \bibinfo{author}{\bibfnamefont{K.}~\bibnamefont{Kohri}},
  \bibinfo{author}{\bibfnamefont{Y.}~\bibnamefont{Sendouda}}, \bibnamefont{and}
  \bibinfo{author}{\bibfnamefont{J.}~\bibnamefont{Yokoyama}}
  (\bibinfo{year}{2020}), \eprint{2002.12778}.

\bibitem[{\citenamefont{K\"uhnel and Freese}(2017)}]{Kuhnel}
\bibinfo{author}{\bibfnamefont{F.}~\bibnamefont{K\"uhnel}} \bibnamefont{and}
  \bibinfo{author}{\bibfnamefont{K.}~\bibnamefont{Freese}},
  \bibinfo{journal}{Phys. Rev. D} \textbf{\bibinfo{volume}{95}},
  \bibinfo{pages}{083508} (\bibinfo{year}{2017}), \eprint{1701.07223}.

\bibitem[{\citenamefont{Niemeyer and Jedamzik}(1998)}]{Niemeyer}
\bibinfo{author}{\bibfnamefont{J.~C.} \bibnamefont{Niemeyer}} \bibnamefont{and}
  \bibinfo{author}{\bibfnamefont{K.}~\bibnamefont{Jedamzik}},
  \bibinfo{journal}{Phys. Rev. Lett.} \textbf{\bibinfo{volume}{80}},
  \bibinfo{pages}{5481} (\bibinfo{year}{1998}), \eprint{astro-ph/9709072}.

\bibitem[{\citenamefont{Green}(2016)}]{Green}
\bibinfo{author}{\bibfnamefont{A.~M.} \bibnamefont{Green}},
  \bibinfo{journal}{Phys. Rev. D} \textbf{\bibinfo{volume}{94}},
  \bibinfo{pages}{063530} (\bibinfo{year}{2016}), \eprint{1609.01143}.

\bibitem[{\citenamefont{Kannike et~al.}(2017)\citenamefont{Kannike, Marzola,
  Raidal, and Veerm\"ae}}]{Kannike}
\bibinfo{author}{\bibfnamefont{K.}~\bibnamefont{Kannike}},
  \bibinfo{author}{\bibfnamefont{L.}~\bibnamefont{Marzola}},
  \bibinfo{author}{\bibfnamefont{M.}~\bibnamefont{Raidal}}, \bibnamefont{and}
  \bibinfo{author}{\bibfnamefont{H.}~\bibnamefont{Veerm\"ae}},
  \bibinfo{journal}{JCAP} \textbf{\bibinfo{volume}{09}}, \bibinfo{pages}{020}
  (\bibinfo{year}{2017}), \eprint{1705.06225}.

\bibitem[{\citenamefont{{Dolgov} and {Silk}}(1993)}]{Dolgov}
\bibinfo{author}{\bibfnamefont{A.}~\bibnamefont{{Dolgov}}} \bibnamefont{and}
  \bibinfo{author}{\bibfnamefont{J.}~\bibnamefont{{Silk}}},
  \bibinfo{journal}{\prd} \textbf{\bibinfo{volume}{47}}, \bibinfo{pages}{4244}
  (\bibinfo{year}{1993}).

\bibitem[{\citenamefont{Bird et~al.}(2016)\citenamefont{Bird, Cholis, Mu\~noz,
  Ali-Ha\"\i{}moud, Kamionkowski, Kovetz, Raccanelli, and Riess}}]{Bird}
\bibinfo{author}{\bibfnamefont{S.}~\bibnamefont{Bird}},
  \bibinfo{author}{\bibfnamefont{I.}~\bibnamefont{Cholis}},
  \bibinfo{author}{\bibfnamefont{J.~B.} \bibnamefont{Mu\~noz}},
  \bibinfo{author}{\bibfnamefont{Y.}~\bibnamefont{Ali-Ha\"\i{}moud}},
  \bibinfo{author}{\bibfnamefont{M.}~\bibnamefont{Kamionkowski}},
  \bibinfo{author}{\bibfnamefont{E.~D.} \bibnamefont{Kovetz}},
  \bibinfo{author}{\bibfnamefont{A.}~\bibnamefont{Raccanelli}},
  \bibnamefont{and} \bibinfo{author}{\bibfnamefont{A.~G.} \bibnamefont{Riess}},
  \bibinfo{journal}{Phys. Rev. Lett.} \textbf{\bibinfo{volume}{116}},
  \bibinfo{pages}{201301} (\bibinfo{year}{2016}), \eprint{1603.00464}.

\bibitem[{\citenamefont{Quinlan and Shapiro}(1989)}]{cross1}
\bibinfo{author}{\bibfnamefont{G.~D.} \bibnamefont{Quinlan}} \bibnamefont{and}
  \bibinfo{author}{\bibfnamefont{S.~L.} \bibnamefont{Shapiro}},
  \bibinfo{journal}{The Astrophysical Journal} \textbf{\bibinfo{volume}{343}},
  \bibinfo{pages}{725} (\bibinfo{year}{1989}).

\bibitem[{\citenamefont{Mouri and Taniguchi}(2002)}]{cross2}
\bibinfo{author}{\bibfnamefont{H.}~\bibnamefont{Mouri}} \bibnamefont{and}
  \bibinfo{author}{\bibfnamefont{Y.}~\bibnamefont{Taniguchi}},
  \bibinfo{journal}{The Astrophysical Journal Letters}
  \textbf{\bibinfo{volume}{566}}, \bibinfo{pages}{L17} (\bibinfo{year}{2002}).

\bibitem[{\citenamefont{O'Leary et~al.}(2009)\citenamefont{O'Leary, Kocsis, and
  Loeb}}]{OLeary}
\bibinfo{author}{\bibfnamefont{R.~M.} \bibnamefont{O'Leary}},
  \bibinfo{author}{\bibfnamefont{B.}~\bibnamefont{Kocsis}}, \bibnamefont{and}
  \bibinfo{author}{\bibfnamefont{A.}~\bibnamefont{Loeb}},
  \bibinfo{journal}{Mon. Not. Roy. Astron. Soc.}
  \textbf{\bibinfo{volume}{395}}, \bibinfo{pages}{2127} (\bibinfo{year}{2009}),
  \eprint{0807.2638}.

\bibitem[{\citenamefont{Press and Schechter}(1974)}]{Press}
\bibinfo{author}{\bibfnamefont{W.~H.} \bibnamefont{Press}} \bibnamefont{and}
  \bibinfo{author}{\bibfnamefont{P.}~\bibnamefont{Schechter}},
  \bibinfo{journal}{Astrophys. J.} \textbf{\bibinfo{volume}{187}},
  \bibinfo{pages}{425} (\bibinfo{year}{1974}).

\bibitem[{\citenamefont{Ludlow et~al.}(2016)\citenamefont{Ludlow, Bose, Angulo,
  Wang, Hellwing, Navarro, Cole, and Frenk}}]{Ludlow}
\bibinfo{author}{\bibfnamefont{A.~D.} \bibnamefont{Ludlow}},
  \bibinfo{author}{\bibfnamefont{S.}~\bibnamefont{Bose}},
  \bibinfo{author}{\bibfnamefont{R.~E.} \bibnamefont{Angulo}},
  \bibinfo{author}{\bibfnamefont{L.}~\bibnamefont{Wang}},
  \bibinfo{author}{\bibfnamefont{W.~A.} \bibnamefont{Hellwing}},
  \bibinfo{author}{\bibfnamefont{J.~F.} \bibnamefont{Navarro}},
  \bibinfo{author}{\bibfnamefont{S.}~\bibnamefont{Cole}}, \bibnamefont{and}
  \bibinfo{author}{\bibfnamefont{C.~S.} \bibnamefont{Frenk}},
  \bibinfo{journal}{Mon. Not. Roy. Astron. Soc.}
  \textbf{\bibinfo{volume}{460}}, \bibinfo{pages}{1214} (\bibinfo{year}{2016}),
  \eprint{1601.02624}.

\bibitem[{\citenamefont{{Loeb} et~al.}(2008)\citenamefont{{Loeb}, {Ferrara},
  and {Ellis}}}]{Loeb}
\bibinfo{author}{\bibfnamefont{A.}~\bibnamefont{{Loeb}}},
  \bibinfo{author}{\bibfnamefont{A.}~\bibnamefont{{Ferrara}}},
  \bibnamefont{and} \bibinfo{author}{\bibfnamefont{R.~S.}
  \bibnamefont{{Ellis}}}, \emph{\bibinfo{title}{{First Light in the Universe}}}
  (\bibinfo{year}{2008}).

\bibitem[{pyc()}]{pyc}
\bibinfo{howpublished}{\url{https://pycbc.org/}}.

\bibitem[{\citenamefont{Zackrisson and Riehm}(2007)}]{Zackrisson}
\bibinfo{author}{\bibfnamefont{E.}~\bibnamefont{Zackrisson}} \bibnamefont{and}
  \bibinfo{author}{\bibfnamefont{T.}~\bibnamefont{Riehm}},
  \bibinfo{journal}{Astron. Astrophys.} \textbf{\bibinfo{volume}{475}},
  \bibinfo{pages}{453} (\bibinfo{year}{2007}), \eprint{0709.1571}.

\bibitem[{\citenamefont{De~Paolis et~al.}(2001)\citenamefont{De~Paolis,
  Ingrosso, and Nucita}}]{De}
\bibinfo{author}{\bibfnamefont{F.}~\bibnamefont{De~Paolis}},
  \bibinfo{author}{\bibfnamefont{G.}~\bibnamefont{Ingrosso}}, \bibnamefont{and}
  \bibinfo{author}{\bibfnamefont{A.}~\bibnamefont{Nucita}},
  \bibinfo{journal}{Astron. Astrophys.} \textbf{\bibinfo{volume}{366}},
  \bibinfo{pages}{1065} (\bibinfo{year}{2001}), \eprint{astro-ph/0011563}.

\bibitem[{\citenamefont{Takahashi and Nakamura}(2003)}]{Takahashi}
\bibinfo{author}{\bibfnamefont{R.}~\bibnamefont{Takahashi}} \bibnamefont{and}
  \bibinfo{author}{\bibfnamefont{T.}~\bibnamefont{Nakamura}},
  \bibinfo{journal}{Astrophys. J.} \textbf{\bibinfo{volume}{595}},
  \bibinfo{pages}{1039} (\bibinfo{year}{2003}), \eprint{astro-ph/0305055}.

\bibitem[{\citenamefont{{Schneider} et~al.}(1992)\citenamefont{{Schneider},
  {Ehlers}, and {Falco}}}]{Schneider}
\bibinfo{author}{\bibfnamefont{P.}~\bibnamefont{{Schneider}}},
  \bibinfo{author}{\bibfnamefont{J.}~\bibnamefont{{Ehlers}}}, \bibnamefont{and}
  \bibinfo{author}{\bibfnamefont{E.~E.} \bibnamefont{{Falco}}},
  \emph{\bibinfo{title}{{Gravitational Lenses}}} (\bibinfo{year}{1992}).

\bibitem[{\citenamefont{Biwer et~al.}(2019)\citenamefont{Biwer, Capano, De,
  Cabero, Brown, Nitz, and Raymond}}]{Biwer:2018osg}
\bibinfo{author}{\bibfnamefont{C.}~\bibnamefont{Biwer}},
  \bibinfo{author}{\bibfnamefont{C.~D.} \bibnamefont{Capano}},
  \bibinfo{author}{\bibfnamefont{S.}~\bibnamefont{De}},
  \bibinfo{author}{\bibfnamefont{M.}~\bibnamefont{Cabero}},
  \bibinfo{author}{\bibfnamefont{D.~A.} \bibnamefont{Brown}},
  \bibinfo{author}{\bibfnamefont{A.~H.} \bibnamefont{Nitz}}, \bibnamefont{and}
  \bibinfo{author}{\bibfnamefont{V.}~\bibnamefont{Raymond}},
  \bibinfo{journal}{Publ. Astron. Soc. Pac.} \textbf{\bibinfo{volume}{131}},
  \bibinfo{pages}{024503} (\bibinfo{year}{2019}), \eprint{1807.10312}.

\bibitem[{\citenamefont{Speagle}(2020)}]{10.1093/mnras/staa278}
\bibinfo{author}{\bibfnamefont{J.~S.} \bibnamefont{Speagle}},
  \bibinfo{journal}{Monthly Notices of the Royal Astronomical Society}
  \textbf{\bibinfo{volume}{493}}, \bibinfo{pages}{3132} (\bibinfo{year}{2020}).

\bibitem[{\citenamefont{Abbott et~al.}(2016{\natexlab{b}})}]{Martynov:2016fzi}
\bibinfo{author}{\bibfnamefont{B.~P.} \bibnamefont{Abbott}}
  \bibnamefont{et~al.}, \bibinfo{journal}{Phys. Rev. D}
  \textbf{\bibinfo{volume}{93}}, \bibinfo{pages}{112004}
  (\bibinfo{year}{2016}{\natexlab{b}}), \bibinfo{note}{[Addendum: Phys.Rev.D
  97, 059901 (2018)]}, \eprint{1604.00439}.

\bibitem[{\citenamefont{Haris et~al.}(2018)\citenamefont{Haris, Mehta, Kumar,
  Venumadhav, and Ajith}}]{india}
\bibinfo{author}{\bibfnamefont{K.}~\bibnamefont{Haris}},
  \bibinfo{author}{\bibfnamefont{A.~K.} \bibnamefont{Mehta}},
  \bibinfo{author}{\bibfnamefont{S.}~\bibnamefont{Kumar}},
  \bibinfo{author}{\bibfnamefont{T.}~\bibnamefont{Venumadhav}},
  \bibnamefont{and} \bibinfo{author}{\bibfnamefont{P.}~\bibnamefont{Ajith}}
  (\bibinfo{year}{2018}), \eprint{1807.07062}.

\bibitem[{\citenamefont{{Belczynski} et~al.}(2010)\citenamefont{{Belczynski},
  {Bulik}, {Fryer}, {Ruiter}, {Valsecchi}, {Vink}, and
  {Hurley}}}]{2010ApJ...714.1217B}
\bibinfo{author}{\bibfnamefont{K.}~\bibnamefont{{Belczynski}}},
  \bibinfo{author}{\bibfnamefont{T.}~\bibnamefont{{Bulik}}},
  \bibinfo{author}{\bibfnamefont{C.~L.} \bibnamefont{{Fryer}}},
  \bibinfo{author}{\bibfnamefont{A.}~\bibnamefont{{Ruiter}}},
  \bibinfo{author}{\bibfnamefont{F.}~\bibnamefont{{Valsecchi}}},
  \bibinfo{author}{\bibfnamefont{J.~S.} \bibnamefont{{Vink}}},
  \bibnamefont{and} \bibinfo{author}{\bibfnamefont{J.~R.}
  \bibnamefont{{Hurley}}}, \bibinfo{journal}{\apj}
  \textbf{\bibinfo{volume}{714}}, \bibinfo{pages}{1217} (\bibinfo{year}{2010}),
  \eprint{0904.2784}.

\bibitem[{\citenamefont{{Spera} et~al.}(2015)\citenamefont{{Spera}, {Mapelli},
  and {Bressan}}}]{2015MNRAS.451.4086S}
\bibinfo{author}{\bibfnamefont{M.}~\bibnamefont{{Spera}}},
  \bibinfo{author}{\bibfnamefont{M.}~\bibnamefont{{Mapelli}}},
  \bibnamefont{and}
  \bibinfo{author}{\bibfnamefont{A.}~\bibnamefont{{Bressan}}},
  \bibinfo{journal}{Monthly Notices of the Royal Astronomical Society}
  \textbf{\bibinfo{volume}{451}}, \bibinfo{pages}{4086} (\bibinfo{year}{2015}),
  \eprint{1505.05201}.

\bibitem[{\citenamefont{Sasaki et~al.}(2018)\citenamefont{Sasaki, Suyama,
  Tanaka, and Yokoyama}}]{Sasaki}
\bibinfo{author}{\bibfnamefont{M.}~\bibnamefont{Sasaki}},
  \bibinfo{author}{\bibfnamefont{T.}~\bibnamefont{Suyama}},
  \bibinfo{author}{\bibfnamefont{T.}~\bibnamefont{Tanaka}}, \bibnamefont{and}
  \bibinfo{author}{\bibfnamefont{S.}~\bibnamefont{Yokoyama}},
  \bibinfo{journal}{Class. Quant. Grav.} \textbf{\bibinfo{volume}{35}},
  \bibinfo{pages}{063001} (\bibinfo{year}{2018}), \eprint{1801.05235}.

\bibitem[{\citenamefont{Abbott et~al.}(2018)}]{Abbottpbh}
\bibinfo{author}{\bibfnamefont{B.}~\bibnamefont{Abbott}} \bibnamefont{et~al.}
  (\bibinfo{collaboration}{LIGO Scientific, Virgo}), \bibinfo{journal}{Phys.
  Rev. Lett.} \textbf{\bibinfo{volume}{121}}, \bibinfo{pages}{231103}
  (\bibinfo{year}{2018}), \eprint{1808.04771}.

\bibitem[{\citenamefont{Abbott et~al.}(2019{\natexlab{b}})}]{Abbottpbh2}
\bibinfo{author}{\bibfnamefont{B.}~\bibnamefont{Abbott}} \bibnamefont{et~al.}
  (\bibinfo{collaboration}{LIGO Scientific, Virgo}), \bibinfo{journal}{Phys.
  Rev. Lett.} \textbf{\bibinfo{volume}{123}}, \bibinfo{pages}{161102}
  (\bibinfo{year}{2019}{\natexlab{b}}), \eprint{1904.08976}.

\bibitem[{Hu()}]{Hu}
\bibinfo{howpublished}{\url{http://background.uchicago.edu/}}.

\end{thebibliography}
%\bibliographystyle{apsrev}
%\bibliography{references1}

\end{document}